\documentclass[aps,prb,amssymb,twocolumn,showpacs]{revtex4}

\usepackage{graphicx}

\newcommand{\bea}{\begin{eqnarray}}

\newcommand{\eea}{\end{eqnarray}}

\newcommand{\beq}{\begin{equation}}

\newcommand{\eeq}{\end{equation}}

\newcommand{\lav}{\langle}

\newcommand{\rav}{\rangle}

\newlength{\textwidthm}

\begin{document}
\def \tr{{\mbox{tr~}}}
\def \ra{{\rightarrow}}
\def \ua{{\uparrow}}
\def \da{{\downarrow}}
\def \be{\begin{equation}}
\def \ee{\end{equation}}
\def \ba{\begin{array}}
\def \ea{\end{array}}
\def \bea{\begin{eqnarray}}
\def \eea{\end{eqnarray}}
\def \nn{\nonumber}
\def \l{\left}
\def \r{\right}
\def \half{{1\over 2}}
\def \etal{{\it {et al}}}
\def \cH{{\cal{H}}}
\def \cM{{\cal{M}}}
\def \cN{{\cal{N}}}
\def \cQ{{\cal Q}}
\def \cI{{\cal I}}
\def \cV{{\cal V}}
\def \cG{{\cal G}}
\def \cF{{\cal F}}
\def \cZ{{\cal Z}}
\def \bS{{\bf S}}
\def \bI{{\bf I}}
\def \bL{{\bf L}}
\def \bG{{\bf G}}
\def \bQ{{\bf Q}}
\def \bK{{\bf K}}
\def \bR{{\bf R}}
\def \br{{\bf r}}
\def \bu{{\bf u}}
\def \bq{{\bf q}}
\def \bk{{\bf k}}
\def \bz{{\bf z}}
\def \bx{{\bf x}}
\def \bpsi{{\bar{\psi}}}
\def \tJ{{\tilde{J}}}
\def \W{{\Omega}}
\def \e{{\epsilon}}
\def \lam{{\lambda}}
\def \L{{\Lambda}}
\def \a{{\alpha}}
\def \t{{\theta}}
\def \b{{\beta}}
\def \g{{\gamma}}
\def \D{{\Delta}}
\def \d{{\delta}}
\def \w{{\omega}}
\def \s{{\sigma}}
\def \f{{\varphi}}
\def \x{{\chi}}
\def \e{{\epsilon}}
\def \h{{\eta}}
\def \G{{\Gamma}}
\def \z{{\zeta}}
\def \hatt{{\hat{\t}}}
\def \hn{{\bar{n}}}
\def \vk{{\bf{k}}}
\def \vq{{\bf{q}}}
\def \gk{{\g_{\vk}}}
\def \nd{{^{\vphantom{\dagger}}}}
\def \yd{^\dagger}
\def \av#1{{\langle#1\rangle}}
\def \ket#1{{\,|\,#1\,\rangle\,}}
\def \bra#1{{\,\langle\,#1\,|\,}}
\def \braket#1#2{{\,\langle\,#1\,|\,#2\,\rangle\,}}

\title{Commensurate mixtures of ultra-cold atoms in one dimension}

\author{L. Mathey}

\affiliation{Physics Department, Harvard University, Cambridge, MA 02138}

\date{\today}

\begin{abstract}
We study binary mixtures
 of ultra-cold atoms, confined
 to one dimension in an optical lattice, with commensurate
 densities.
Within a Luttinger liquid description, which treats various mixtures
 on equal footing, 
  we derive a system 
 of renormalization group equations at second order, 
  from which we determine the rich phase diagrams
 of these mixtures.
These phases include charge/spin density wave order,
 singlet and triplet pairing, polaron pairing\cite{mathey}, 
and a supersolid phase.
 Various methods to detect our results experimentally are discussed.
\end{abstract}


\maketitle


%
%
%
%
%
%
%
%

\section{Introduction}\label{intro}
%
%
%
Recent advances in controlling ultra cold atoms
 lead to  the realization
of truly one dimensional systems, and the
 study of many-body effects therein. Important benchmarks,
such as the Tonks-Girardeau gas \cite{paredes,weiss} and
the Mott transition in one dimension\cite{stoeferle}, have been achieved
by trapping bosonic atoms in tight tubes formed by an optical
lattice potential. Novel transport properties of
one dimensional lattice bosons have been studied using these
techniques\cite{fertig}.
More recently, a strongly interacting one dimensional Fermi gas
was realized using similar trapping methods\cite{Moritz}.
Interactions between the fermion atoms
were
controlled
by tuning a Feshbach resonance
in these experiments.
On the theory side, numerous proposals were given for realizing 
a variety of different phases in ultra cold Fermi systems
\cite{theory1, theory2},
 Bose-Fermi mixtures\cite{cazalilla,mathey,mathey2,pinaki}, as well as 
 Bose-Bose mixtures\cite{isacsson}.
 In [\onlinecite{kuklov}], 
 commensurate mixtures in higher dimensions were studied.

In this paper we explore the behavior of ultracold atomic mixtures, confined 
to one-dimensional (1D) 
motion in an optical lattice, that exhibit different types of 
commensurability, by which
 we mean that the atomic densities and/or the inverse 
 lattice spacing have an integer ratio.
Commensurable fillings arise naturally in many ultracold atom systems, because 
the external trap potential approximately corresponds to a sweep of the 
chemical potential through the phase diagram, and therefore passes through points 
of commensurability.
At these
points the system can
 develop an energy gap, which fixes
 the density commensurability over a spatially extended volume.
%
%
%
%
%
%
%
%
%
%
This was demonstrated in 
 the celebrated Mott insulator 
experiment by Greiner et al.\cite{greiner}, 
 where Mott phases with integer filling 
 occurred in shell-shaped regions in the atom trap.
These gapped phases gave rise to the  
 well-known signature in the time-of-flight images\cite{trap},
 and triggered the endeavor of `engineering' many-body states
 in optical lattices.
Further examples include the recently created density-imbalanced
 fermion mixtures \cite{fmixtures} in which the development of
 a balanced, i.e. commensurate,
  mixture at the center of the trap is observed.

In 1D, this phenomenon is of particular importance,
 because it is the only effect that can lead to the opening of
 a gap, for a system with short-range interactions. 
 In contrast to higher dimensional systems, where, for instance, pairing
 can lead to a state with an energy gap,
 in 1D  only discrete symmetries
 can be broken, due to the importance of fluctuations.
Orders that correspond to a
 continuous symmetry can, at most, develop quasi long range order (QLRO),
 which refers to a state in which an order parameter $O(x)$
 has a correlation function with algebraic scaling, 
$\lav O(x) O(0)\rav \sim |x|^{-(2-\alpha)}$, with
 a positive scaling exponent $\alpha$.

Due to its importance in solid state physics, the most thoroughly
 studied commensurate 1D system is the SU(2) symmetric
 system of spin-1/2 fermions.
 This system develops a spin gap for attractive 
 interaction and remains gapless 
 for repulsive interaction, as
 can be seen from a second order RG calculation.
 However, the assumed symmetry between the
 two internal spin states, which is natural in solid state systems,
 does not generically occur 
in Fermi-Fermi mixtures (FFMs) 
of ultra-cold atoms, where
 the `spin' states are in fact different 
 hyperfine states of the atoms. 
 An analysis of the generic system 
 is therefore highly called for.
 Furthermore, we will extend this analysis
 to both Bose-Fermi (BFMs) and Bose-Bose mixtures (BBMs), as well
 as to the dual commensurability, in which the charge field, and not
 the spin field, exhibits commensurate filling, as will be explained
 below.

\begin{figure}
\includegraphics[width=7cm]{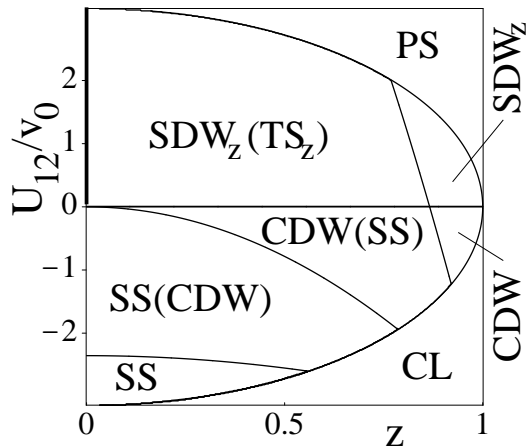}
\caption{\label{FFMfig} 
Phase diagram of a commensurate FFM or a
 BBM of
 hardcore bosons (with the replacement $TS_z\ra SS$), 
in terms of the interaction $U_{12}$
and the parameter $z=|v_1-v_2|/(v_1+v_2)$.
For both attractive  and repulsive 
 interactions a spin gap opens, except for
 $z=0$ and positive interaction. 
In the attractive regime, a FFM or a BBM shows
 either singlet pairing or CDW order, or a coexistence of these phases. 
 For repulsive interaction
 these mixtures show SDW ordering, with FFMs and BBMs
 showing subdominant triplet or singlet pairing, respectively,
  for a large range of $z$.
In the gapless
 regime, a FFM shows degenerate 
SDW and CDW order, 
 and a BBM shows SF with subdominant 
 CDW, i.e. supersolid behavior.
For very large positive values of $U_{12}$ the system undergoes
phase separation (PS); for very large negative values it collapses (CL).
}
\end{figure}

The main results of this paper are the phase
 diagrams shown in Fig. \ref{FFMfig}--\ref{BFM2fig}.
 We find that both attractive and repulsive interactions
 can open an energy gap.
 For FFMs the entire phase diagram is gapped, except for
 the repulsive SU(2) symmetric regime (cp. [\onlinecite{theory2}]), for
 BFMs or BBMs the bosonic liquid(s) need(s) to be close to the
 hardcore limit, otherwise the system remains gapless.
Furthermore, we find a rich structure of quasi-phases,
 including charge and spin density wave order (CDW, SDW), 
 singlet and triplet pairing (SS, TS), polaron pairing \cite{mathey,mathey2},
 and a supersolid phase, which is the first example 
 of a supersolid phase in 1D.
These results are derived within a Luttinger liquid (LL) 
 description, which treats bosonic and fermionic liquids on equal footing.
%
%
%
%
%
%

%

This paper is organized as follows: In Section \ref{class} we classify the
 different types of commensurate mixtures that can occur, 
 and in Section \ref{effact} we discuss the effective action of the
 mixtures with the most relevant commensurability term.
 In Section \ref{RG} we discuss the set of renormalization group
 equations for such systems, and in Section \ref{FFM}, \ref{BBM}, and \ref{BFM},
  we apply these results to Fermi-Fermi, Bose-Bose, and Bose-Fermi mixtures,
 respectively. In Section \ref{detect}, we discuss the experimental
 detectibility, and in Section \ref{conclude} we conclude.

\section{Classification of commensurate mixtures}\label{class}
We will now classify the
 types of commensurability that can occur in
 a system with short-ranged density-density
 interaction.
 We consider Haldane's representation \cite{Haldane}
of the densities for the two species:
\bea
n_{1/2} & = & [\nu_{1/2} + \Pi_{1/2}] \sum_{m} e^{2 m i\Theta_{1/2}}
\eea
$\nu_1$ and $\nu_2$ are the densities of the two liquids,
$\Pi_{1/2}(x)$ are the low-k parts (i.e. $k\ll 1/\nu$) of the density fluctuations; 
 the fields  
$\Theta_{1/2}(x)$ are given by 
 $\Theta_{1/2}(x) 
= \pi \nu_{1/2} x + \theta_{1/2}(x)$, 
with $\theta_{1/2}(x)=\pi \int^x dy \Pi_{1/2}(y)$.
%
%
%
 These expressions hold for both bosons and fermions.
If we use this representation in a density-density 
 interaction 
term $U_{12}\int dx n_1(x)n_2(x)$,
we generate to lowest order a term of the shape $U_{12}\int dx \Pi_1(x)\Pi_2(x)$, 
 but in addition an infinite number of nonlinear terms, corresponding
to all harmonics in the representation.
However, only the terms for which the linear terms ($2 \pi m_{1/2} \nu_{1/2}x$) cancel,
can drive a phase transition.
For a continuous system this happens for $m_1 \nu_1 - m_2\nu_2=0$,
whereas for a system on a lattice we have the condition $m_1 \nu_1 - m_2\nu_2=m_3$, 
where $m_1$,$m_2$ and $m_3$ are integer numbers.
%
%
In general, higher integer numbers correspond to terms
 that are less relevant, because the scaling dimension
 of the non-linear term scales quadratically with these integers.
 We are therefore lead to consider
 small integer ratios between
 the fillings and/or the lattice if present.
In [\onlinecite{mathey2}], we considered
 two cases of commensurabilities: a Mott insulator transition
 coupled to an incommensurate liquid,
 and a fermionic liquid at
 half-filling coupled to an incommensurate bosonic liquid. 
%
In both cases
 the
 commensurability occurs between  one species and the
 lattice, but does not involve the second species.
In this
 paper we 
 consider the 
 two most relevant, i.e. lowest order,  
cases which exhibit a commensurability that 
 involves both species. 
 The first case is the case 
 of equal filling $\nu_1=\nu_2$, 
 the second is the case 
 of the total density being unity, i.e. $\nu_1+\nu_2=1$,
 where the densities $\nu_{1}$ and $\nu_2$ themselves are incommensurate.
The first case can drive the system to a spin-gapped state, the
 second to a charge gapped state.
 We will determine in which parameter
 regime these transitions occur,
 and what type of QRLO the
 system exhibits in the
 vicinity of the transition. 
These two cases can be mapped onto each other via a dual mapping, which
enables us to study only one
 case and then infer the results for the
 second by using this mapping.
We will 
write out our discussion
 for the case of equal filling
 and merely state 
the corresponding results for 
 complementary filling.

\section{Effective action}\label{effact}
The action of a two-species mixture
with equal filling
in bosonized form is given by:
\bea\label{S}
S & = & S_{0,1} + S_{0,2} + S_{12} +S_{int}.
\eea
The terms $S_{0,j}$, with $j=1,2$, are given by
\bea
S_{0, j} & = & \frac{1}{2\pi K_{j}} \int d^2r \Big( \frac{1}{v_{j}}(\partial_{\tau} \theta_{j})^2 +  v_{j} (\partial_x \theta_{j})^2 \Big)
\eea
Each of the two types of atoms, regardless of being bosonic or fermionic,
are characterized by a Luttinger parameter $K_{1/2}$ and a velocity $v_{1/2}$.
Here we integrate over ${\bf r}=(v_0 \tau, x)$, 
 where we defined the
 energy scale $v_0=(v_1+v_2)/2$.
%
%
%
%
The term $S_{12}$ describes the acoustic coupling between the two species, and is bilinear:
\bea
S_{12} & = &  \frac{U_{12}}{\pi^2} \int d^2r \partial_x \theta_1 \partial_x \theta_2 
 + \frac{V_{12}}{\pi^2} \int d^2r \partial_\tau \theta_1 \partial_\tau \theta_2.
\eea
The second term is created during the RG flow; its prefactor therefore has the initial value
$V_{12}(0)=0$.
We define $S_0 =  S_{0,1} + S_{0,2} + S_{12}$, which is
 the diagonalizable part of the action. 
%
%
%
$S_{int}$ 
 corresponds 
to the non-linear coupling between the two liquids,
which we study within an RG approach:
\bea
S_{int} & = &  \frac{2 g_{12}}{(2 \pi \alpha)^2} \int d^2r  
\cos(2 \theta_1 - 2\theta_2).
\label{Sint}
\eea
%
%
%
%
%
%
%
%
This 
 bosonized 
description 
  applies to a BBM, a
 BFM, and a FFM.
Depending on which of these mixtures 
we want to describe we either 
construct bosonic or fermionic operators 
according to Haldane's contruction \cite{Haldane}:
\bea
f/b & = & [\nu_0 + \Pi]^{1/2} \sum_{m\, odd / even} e^{m i\Theta} e^{i\Phi}.
\eea
$\nu_0$ is the zero-mode of the density,
$\Phi(x)$ is the phase field, which is the conjugate field of the density fluctuations $\Pi(x)$.
The action for a mixture with complementary
 filling, $\nu_1+\nu_2 =1$, is of the form $S_0+S'_{int}$, where
 the interaction $S'_{int}$ is given by:
\bea
S'_{int} & = &  \frac{2 g_{12}}{(2 \pi \alpha)^2} \int d^2r  
\cos(2 \theta_1 + 2\theta_2).
\label{Sint'}
\eea
To map the action in Eq. (\ref{S}) onto
 this system we use
 the mapping: $\theta_2\ra -\theta_2$, $\phi_2\ra -\phi_2$, and 
$g_{12}\ra -g_{12}$, which evidently
 maps a mixture with complementary filling and attractive (repulsive)
 interaction and onto a mixture with equal
 filling with repulsive (attractive) interaction.

\section{Renormalization group}\label{RG}
 To study the action given in Eq. (\ref{S}), 
we perform an RG calculation along the lines of the
treatment of the sine-Gordon model in [\onlinecite{SG}].
In our model, 
 a crucial modification
 arises: the linear combination  
$\theta_1 - \theta_2$, that appears
 in the non-linear term,  is not proportional to an eigenmode
of $S_0$, and therefore the RG flow does not affect only  one
 separate sector of the system, as in an
SU(2)-symmetric system.
%
The RG scheme that we use here
 proceeds as follows:
First, we diagonalize $S_0$ through the transformation  
\bea
\label{diag1}
\theta_1 & = & B_1 \tilde{\theta}_1 + B_2\tilde{\theta}_{2},\\  
\theta_2 & = & D_1 \tilde{\theta}_1 + D_2\tilde{\theta}_{2}.\label{diag2}
%
\eea
%
%
%
%
%
%
%
%
%
%
%
%
%
%
%
%
%
 The coefficients 
 $B_{1/2}$ and
 $D_{1/2}$ are given in the Appendix.
%
%
%
The fields  
$\tilde{\theta}_{1/2}$ are the eigenmode
 fields with velocities
  $\tilde{v}_{1/2}$ (see Appendix).
%
%
%
%
%
%
%
%
%
%
%
%
%
\begin{figure}
\includegraphics[width=7cm]{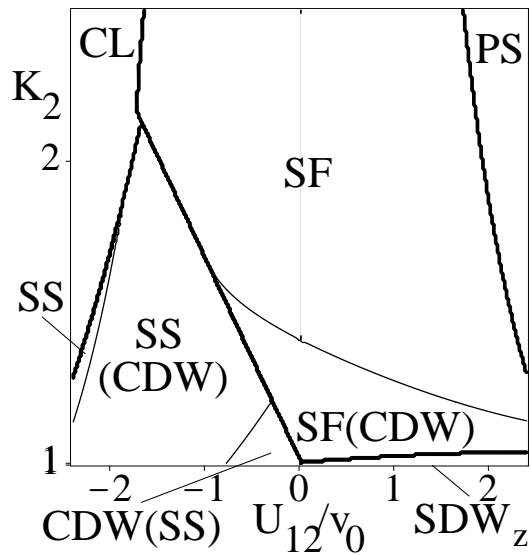}
\caption{\label{BBMfig}
 Phase diagram of a BBM with the first 
species being in the hardcore limit,
 in 
terms of $U_{12}$, and 
the Luttinger parameter of the second species ($K_2$), at
 the fixed 
 velocity ratio $|v_1-v_2|/(v_1+v_2)=0.5$.
For large repulsive interaction the system undergoes phase separation (PS), for large attractive
interaction the system collapses (CL).
In the regime below the thick line the system opens
 a gap, i.e. if species 2 is close to the hardcore limit.
 However, for larger values of $K_2$,  
the gapless phase is restored.
 Close to the
 transition, the properties of the hardcore bosons,
 are affected by the RG flow, leading
   to supersolid behavior.
}
\end{figure}
%
%
%
%
%
%
%
%
%
%
%
 As the next step, 
we introduce an energy cut-off $\Lambda$ on
 the fields $\tilde{\theta}_{1/2}$ 
according to
$\omega^2/\tilde{v}_{1/2} + \tilde{v}_{1/2}k^2 < \Lambda^2$.
We shift this cut-off by an amount $d\Lambda$,
and correct for this shift up to second order in $g_{12}$.
At first order, only $g_{12}$
  is affected, its
flow equation is given by:
\bea
\frac{d g_{12}}{d l} & = & \Big(2 - K_1 - K_2 - 
\frac{2}{\pi}\frac{U_{12} + V_{12} v_1 v_2}{v_1 + v_2}\Big) 
g_{12}, \label{RG_g12}
\eea
%
%
%
 with $dl=d\Lambda/\Lambda$. 
At second order several terms are created that are 
quadratic in the original fields $\theta_1$ and $\theta_2$.
We undo the diagonalization, Eq. (\ref{diag1}) and (\ref{diag2}), and absorb
these terms into the parameters of the action, which
 concludes the RG step.
 By iterating this procedure we obtain these flow equations
 at second order
in $g_{12}$:
\bea
\frac{d K_{1/2}}{d l} & = & - \frac{g_{12}^2}{16\pi^2} 
\Big(2+\Big(\frac{v_2}{v_1}+\frac{v_1}{v_2}\Big)\Big)\label{RG_K1}\\
\frac{d v_1}{d l} & = &   v_1\frac{g_{12}^2}{16\pi^2} 
\Big(\frac{v_2}{v_1}-\frac{v_1}{v_2}\Big)\label{RG_v1}\\
\frac{d v_2}{d l} & = & v_2 \frac{g_{12}^2}{16\pi^2} 
\Big(\frac{v_1}{v_2}-\frac{v_2}{v_1}\Big)\label{RG_v2}\\
\frac{d U_{12}}{d l} & = & - \frac{g_{12}^2}{8\pi} (v_1+v_2)\label{RG_U12}\\
\frac{d V_{12}}{d l} & = & - \frac{g_{12}^2}{8\pi} (1/v_1+1/v_2)\label{RG_V12}
\eea
%
%
%
%
%
%
%
%
%
%
%
%
%
%
%
%
%
A similar set of equations has been
derived in [\onlinecite{theory2}] for a FFM
 in non-bosonized form.
The difference between our result and the result
in [\onlinecite{theory2}]
is the 
renormalization of the velocities,  
that we find 
here, 
which 
is due to different types of expansions:
In [\onlinecite{theory2}] only one-loop contributions
are taken into account, whereas here we use a
cumulant expansion in $g_{12}$, which at second
order includes contributions that are two-loop for the
renormalization of the velocities.
  These contributions, which
would integrate to zero for equal velocities, 
as can be 
seen from Eqns. \ref{RG_v1} and 
\ref{RG_v2}, leads to the discrepancy between the expansion
in the number of loops and the cumulant expansion, 
 and 
 gives
 a small quantitative correction of the velocities.
  As mentioned before, the advantage of the current
 approach is that the QRLO of the 
 system can be directly determined from the
 resulting renormalized parameters, and that
 the same action can be used to study BBMs and BFMs.

The system of differential equations, Eqns. (\ref{RG_g12}) 
to (\ref{RG_V12}),
 can show two types of qualitative behavior:
 The coefficient $g_{12}$ of the non-linear term (\ref{Sint}) can
 either flow to zero, i.e. $S_{int}$ is irrelevant, or it diverges,
 leading to the formation of an energy gap.
 In the first case, the system flows to
 a fixed point that is described
 by a renormalized diagonalizable
  action of the type $S_0$, from which the quasi-phases can be
 determined.

%
%
%

 When $S_{int}$ is relevant,
%
 we 
introduce the fields \cite{review}
%
%
%
$\theta_{\rho/\sigma} = \frac{1}{\sqrt{2}}(\theta_1 \pm \theta_2)$,
%
%
%
which define the charge and the spin sector of the system.
 In this regime, 
 these sectors 
 decouple.
%
Each of the
 two sectors is characterized by a Luttinger parameter
and a velocity, $K_{\rho/\sigma}$ and $v_{\rho/\sigma}$,
 which are related to the original parameters
 in $S_0$ in a
straightforward
 way. 
Using the numerical solution of the flow equations, 
we find that $K_\sigma\rightarrow 0$, as can be expected for
an ordering of the nature of a spin gap, leaving $K_\rho$
the only parameter characterizing the QLRO in this phase.
%
%
%
%
%
%
%
%
%
%
%
%
%
%
%
%
%

In order to determine
 the QLRO in the system we  determine
 the scaling
 exponents of various order parameters.
 For that purpose, we use the 
 bosonization representation of these order parameters,
 which contain the fields $\theta_{1/2}$
 and $\phi_{1/2}$, and use the diagonalization, Eqs. (\ref{diag1}) 
 and (\ref{diag2}), 
 for the fields $\theta_{1/2}$, as well
 as the dual transformation for the fields $\phi_{1/2}$:
\bea\label{dualdiag1}
\phi_1 & = &  C_1 \tilde{\phi}_1 + C_2\tilde{\phi}_{2},\\
\phi_2 & = & E_1 \tilde{\phi}_1 + E_2\tilde{\phi}_{2}.\label{dualdiag2}
\eea
The coefficients $C_{1/2}$ and $D_{1/2}$ are given in the Appendix.
%
%
%
%
%
Since the order parameters are now written in terms
 of the eigenfields $\tilde{\theta}_{1/2}$
 and $\tilde{\phi}_{1/2}$, the correlation functions can be 
 evaluated in a straight forward manner.
 The scaling exponents are given by various quadratic
 expressions of the parameters in Eqs. (\ref{diag1}), (\ref{diag2}),
 (\ref{dualdiag1}), and (\ref{dualdiag2}).
 In [\onlinecite{mathey2}], we give an extensive list
 of correlation functions, which can be
 transferred to the system considered here, with the
 formal replacement: $\beta_{1/2}\rightarrow B_{1/2}$,
 $\gamma_{1/2}\rightarrow C_{1/2}$,  $\delta_{1/2}\rightarrow D_{1/2}$, 
 and  $\epsilon_{1/2}\rightarrow E_{1/2}$.
 The order parameter with the largest positive scaling exponent
 shows the dominant order, whereas other orders with positive
 exponent are subdominant.

\begin{figure}
\includegraphics[width=7cm]{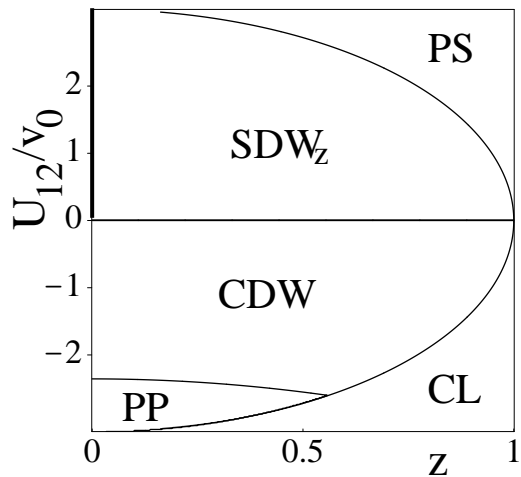}
\caption{\label{BFM1fig}
 Phase diagram of a BFM with hardcore bosons, 
in terms of the interaction $U_{12}$
and the parameter $z=|v_1-v_2|/(v_1+v_2)$.
For both attractive  and repulsive 
 interactions a spin gap opens, except for
 $z=0$ and positive interaction. 
In the attractive regime, 
 a BFM shows either CDW order or polaron pairing; 
 for repulsive interaction
 BFMs show SDW ordering.
In the gapless
 regime, 
 a BFM shows CDW order for the fermions
 and SF for the bosons.
For very large positive values of $U_{12}$ the system undergoes
phase separation (PS); for very large negative values it collapses (CL).
}
\end{figure}
\section{Fermi-Fermi mixtures}\label{FFM}
We will now apply this procedure to the different types of mixtures.
For a FFM 
 we find that the system always develops a
 gap, with the exception of the repulsive SU(2) symmetric regime 
 (cp. [\onlinecite{theory2}]).
 To determine the QLRO 
we introduce the following operators \cite{review, giamarchi}:
\bea
O_{SS} &=&  \sum_{\sigma, \sigma'} \tilde{\sigma} f_{R,\sigma} \delta_{\sigma,\sigma'} 
f_{L, 3-\sigma'},\\ 
O_{TS}^a& =&  \sum_{\sigma, \sigma'} \tilde{\sigma} f_{R,\sigma} \sigma^a_{\sigma,\sigma'}
 f_{L, 3-\sigma'},\\ 
O_{CDW}& =& \sum_{\sigma, \sigma'} f_{R,\sigma}^\dagger \delta_{\sigma,\sigma'} 
f_{L,\sigma'},\\
O_{SDW}^a& =&  \sum_{\sigma, \sigma'} \tilde{\sigma} f_{R,\sigma}^\dagger 
\sigma^a_{\sigma,\sigma'}
 f_{L,\sigma'},
\eea
 with $\sigma, \sigma'=1,2$, $\tilde{\sigma}=3-2\sigma$, and $a=x,y,z$. 
In the gapless SU(2) symmetric regime, 
 both CDW and SDW show QLRO, with both scaling exponents of the
 form $\alpha_{SDW/CDW}=1-K_\rho$\cite{review}, which
 shows that these
 orders are algebraically degenerate.
Within
 the gapped regime
 the scaling exponents of 
these operators are given by $\alpha_{SS,TS_z}= 2 - K_\rho^{-1}$ and
$\alpha_{CDW,SDW_z} = 2 - K_\rho$. 
As discussed in [\onlinecite{giamarchi}], the
 sign of $g_{12}$ determines
 whether CDW or SDW$_z$, and SS or TS$_z$ appears.
 In Fig. \ref{FFMfig}, 
we show the phase diagram based on these results.
 In addition to these phases we indicate the appearance
 of the Wentzel-Bardeen instability, shown as phase separation
 for repulsive
 interaction and collapse for attractive interaction.

We will now use the dual mapping
 to obtain the
 phase diagram of a FFM with complementary
 filling from Fig. \ref{FFMfig}. 
 Under this mapping, the attractive and repulsive regimes
 are exchanged with the following replacements: 
 $CDW\ra SDW_z$, $SDW_z\ra CDW$, $SS,TS_z\ra SDW$, and $SDW\ra SS$.
 Note that the gapless regime
 is now on the attractive side, 
  with degenerate CDW and SS pairing.

%
%
%
%
%
%
%

%
%
\section{Bose-Bose mixtures}\label{BBM}
For BBMs we proceed in the
 same way as for FFMs. 
We introduce the following set of order parameters:
\bea
 O_{CDW}&=&b_1^\dagger b_1 +b^\dagger_2 b_2,\\ 
O_{SS}&=&b_1 b_2,\\
  O_{SDW_z}&=&b_1^\dagger b_1 -b^\dagger_2 b_2,\\ 
  O_{SDW_x}&=&b_1^\dagger b_2 +b^\dagger_2 b_1,\\
  O_{SDW_y}&=&-i (b_1^\dagger b_2 -b^\dagger_2 b_1),
\eea
 and
 in addition the superfluid (SF) order
 parameters $b_1$ and $b_2$.
 In  Fig. \ref{FFMfig}  
 we show the phase
 diagram of a mixture of a BBM of hardcore bosons, which
 is almost identical to the one of a FFM.
The phase diagram of the mixture with complementary
 filling, as obtained from the dual mapping, is
 also of the same form as its fermionic equivalent,
 with the
 exception of the gapless regime, in which BBMs show
 supersolid behavior (coexistence
 of SF and CDW order), and with the replacement $TS_z\ra SS$.

In Fig. \ref{BBMfig}, we  
show the phase diagram of 
 a mixture of hardcore bosons (species 1) and bosons in the intermediate
 to hardcore regime (species 2).
 If species 2 is sufficiently far
 away from the hardcore limit, the system remains gapless.
 However, in the vicinity
 of the transition the scaling exponents of the liquids
 are affected by the RG flow. As indicated,
 the effective scaling exponent
 of the hardcore bosons 
  is renormalized to a value that is smaller than 1, and therefore
 we find both SF and CDW order, i.e. supersolid behavior.
 The phase diagram of the dual mixture is of the following form:
  the attractive and the repulsive regime are exchanged,  
 and in the gapped phase we again have the 
 mapping: 
 $CDW\ra SDW_z$, $SDW_z\ra CDW$, $SS\ra SDW_{x,y}$, and $SDW_{x,y}\ra SS$.
 The gapless regime is unaffected.

The paired SF state discussed in [\onlinecite{kuklov}] corresponds 
 to the SS phase discussed here, whereas the dual SDW$_{x,y}$ phase
 that appears for complementary filling corresponds to
 the super-counter-fluid phase described therein.
  Note that here these orders compete with either CDW or SDW$_z$ order,
  and only appear as QLRO, not LRO, as in higher dimensions.
 Both of these insights can only be gained by the using the LL description
 and RG that is used in this paper.


%
%
%
%
\section{Bose-Fermi mixtures}\label{BFM}
\begin{figure}
\includegraphics[width=7cm]{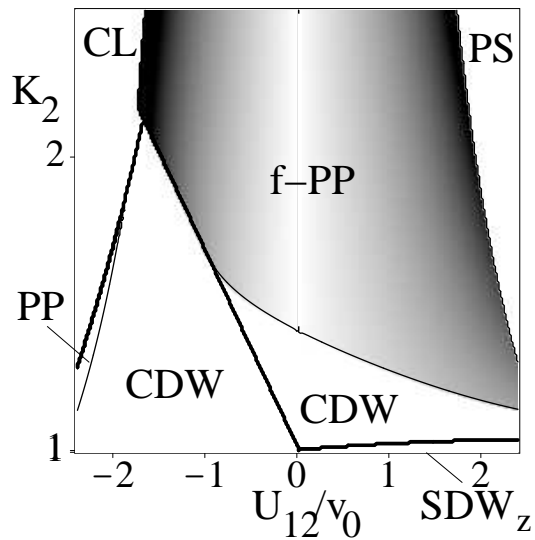}
\caption{\label{BFM2fig}
Phase diagram of a BFM,
 in
terms of $U_{12}$, and 
the Luttinger parameter of the second species ($K_2$), at
 the fixed 
 velocity ratio $|v_1-v_2|/(v_1+v_2)=0.5$.
For large repulsive interaction the system undergoes phase separation (PS), for large attractive
interaction the system collapses (CL).
In the regime below the thick line the system opens
 a gap, i.e. if species 2 is close to the hardcore limit.
 However, for larger values of $K_2$,  
the gapless phase is restored.
 Close to the
 transition, the properties of the fermions,  
 are still affected by the RG flow, leading
 to CDW order.
}
\end{figure}

For a BFM 
 we find that the order parameters $O_{CDW}$, $O_{SDW_z}$, 
 the polaron pairing operator
\bea\label{O_PP}
 O_{f-PP}&=&f_R f_L e^{-2i\lambda \Phi_b}
\eea
(see [\onlinecite{mathey,mathey2}]),
and $b$ can develop QLRO in the gapless regime. 
%
In the gapped regime, the order parameters
\bea
O_{PP}&\equiv& f_R b f_L b,\\
O_{PP'}&\equiv& f_R b^\dagger f_L b^\dagger,
\eea
 in addition to $O_{CDW}$,
show 
  QLRO.
 ($O_{PP/PP'}$ are special cases of the polaron pairing operator
 (\ref{O_PP}), extensively discussed
 in [\onlinecite{mathey}] and [\onlinecite{mathey2}].) 
In Fig. \ref{BFM1fig} we show
 the phase diagram
 of a BFM with hardcore bosons, and
  in Fig. \ref{BFM2fig}, 
 we vary the Luttinger parameter
 of the bosons.
 In both the gapless phase and the gapped phase, we find that
 CDW and $f$-PP or PP, respectively, are mutually exclusive
 and cover the entire phase diagram, cp. [\onlinecite{mathey,mathey2}].
The dual mapping 
 again maps attractive and repulsive regimes onto each other. Within
 the
 gapped phase we find the mapping
 $CDW\ra SDW_z$, $SDW_z\ra CDW$, and $PP\ra PP'$, the 
gapless regime is unaffected.

%
%
%
%

 %

 
\section{Experimental detection}\label{detect}
The phase diagrams that have been derived and 
shown in  Figs. \ref{FFMfig}--\ref{BFM2fig}, are
 given in terms of the parameters
 that appear in the effective action.
 With such a field theoretical approach 
we can find the correct qualitative long-range
behavior, such as the functional form of the correlation functions.
 However, it is also intrinsic to this
 approach that the effective parameters
 appearing can only be qualitatively related to the
  underlying microscopic parameters.
  Based on a phase diagram such as Fig. \ref{BBMfig}, for instance, 
 the following features for 
 a mixture of bosonic atoms with a short-range interaction can be
 expected:
 If one species is in the hardcore limit, and the other 
 is in between an intermediate interaction regime
 and the hardcore limit, then for attractive
 interaction between them, a gapped state can
 be created, in which there is a competition 
 of SS pairing and CDW order. For repulsive interaction, and the
 second species being very close to the hardcore regime, one can also
 expect a gapped phase, in which we find SDW$_z$ order.
 For the intermediate regime we expect a supersolid phase.

Before we conclude,
 we discuss how the
 predictions presented in this paper
 could be measured experimentally.
Since the appearance of a gapped state has already been
 demonstrated for the MI-SF transition in 1D [\onlinecite{stoeferle}],
 and since it constitutes a 
 significant qualitative change in the system, 
 this would be   
 the first feature predicted in this paper to look for.
 As demonstrated in [\onlinecite{RF}], 
 RF spectroscopy can be used
 to determine
 the presence and the size
 of an energy gap.
 To detect the rich structure of QLRO the following approaches
 can be taken: 
CDW order
 will create additional peaks in TOF images, corresponding to
 a wavevector $Q=2k_f$.
 As demonstrated and pointed out in [\onlinecite{noise}],
 the noise in TOF images
 allows to identify
   the different regimes
 of both gapped and gapless phases.
As discussed in [\onlinecite{mathey,mathey2}],
 a laser stirring experiment could determine
 the onset of CDW order for fermions, or the supersolid regime
 for bosons.

\section{Conclusion}\label{conclude}
In conclusion, we have studied mixtures of ultra-cold
 atoms in 1D with commensurate filling.
We used a Luttinger liquid description which enables us to
  study FFMs, BFMs, and BBMs in a single
 approach. 
We find that FFMs are generically gapped for both attractive and repulsive interactions,
 whereas for BFMs and BBMs  the bosons need to be close
 to the hardcore limit.  
 We find a rich structure of quasi-phases in the vicinity 
 of these transitions, in particular a supersolid
 phase for BBMs, that occurs 
  close to the hardcore limit. 
Experimental methods to detect the predictions were also discussed.

We gratefully acknowledge important discussions with 
 D.-W. Wang, A.H. Castro Neto,
 S. Sachdev,  T. Giamarchi, E. Demler, and H.-H. Lin.

\appendix
\section{}\label{App}
Here we give the coefficients that appear in the 
 transformations (\ref{diag1}), (\ref{diag2}), (\ref{dualdiag1}),  
and (\ref{dualdiag2}), that map the original fields on the eigenfields
 at each point in the RG flow.
The coefficients 
 $B_{1/2}$ and
  $D_{1/2}$ are given  
 by: 
\bea
B_1&=&\beta_1 \zeta_1 + \beta_2 \kappa_1,\,  B_2 = \beta_1 \zeta_2 +\beta_2\kappa_2,\\ 
D_1&=& \delta_1 \zeta_1 + \delta_2 \kappa_1, \, 
D_2 = \delta_1 \zeta_2 +\delta_2\kappa_2.
\eea
The coefficients 
 $\beta_{1/2}$ and $\delta_{1/2}$ are given
in [\onlinecite{mathey,mathey2}], 
where the indices 'f' and 'b' need to replaced by
'1' and '2', respectively. 
The other 
coefficients are given by:
\bea
\zeta_1 & = &\sqrt{\tilde{v}_1/v_A}\cos\theta, \,  \zeta_2=\sqrt{\tilde{v}_2/v_A}\sin\theta,\\
\kappa_1 & =& -\sqrt{\tilde{v}_1/v_a}\sin\theta, \, \kappa_2=\sqrt{\tilde{v}_2/v_a}\cos\theta,
\eea
where the angle $\theta$ is given by:
\bea
\tan 2 \theta = -\tilde{V}_{12}/\sqrt{v_A v_a} (\tilde{v}_A^{-2} - \tilde{v}_a^{-2}), 
\eea
where we used:
\bea
\tilde{v}_A & = &v_A/\sqrt{1+2V_{12}v_A\beta_1\delta_1/\pi},\\
\tilde{v}_a&=&v_a/\sqrt{1+2V_{12}v_A\beta_2\delta_2/\pi}.
\eea
 The velocities $\tilde{v}_{1/2}$, corresponding
 to the eigenmodes $\tilde{\theta}_{1/2}$, are given by
\bea
\tilde{v}_{1/2}^{-2} & = & \frac{1}{2}(\tilde{v}_A^{-2}
+\tilde{v}_a^{-2}) \pm \frac{1}{2}\sqrt{(\tilde{v}_A^{-2}
-\tilde{v}_a^{-2})^2 + \tilde{V}_{12}^2 v_A^{-1} 
v_a^{-1}}\nonumber\\
\eea
where $\tilde{V}_{12}$ is given by 
\bea
\tilde{V}_{12}= 2 V_{12} (\beta_1\delta_2+\beta_2\delta_1)/\pi.
\eea
$v_{A,a}$ are defined as in
[\onlinecite{mathey,mathey2}].
The coefficients $C_{1/2}$ and $D_{1/2}$, that appear
 in the dual 
 transformation, Eqs. (\ref{dualdiag1}) and (\ref{dualdiag2}), 
 are given by: 
\bea
C_1&=&\gamma_1 \eta_1 + \gamma_2 \lambda_1, \, C_2 = \gamma_1 \eta_2 +\gamma_2\lambda_2,\\ 
E_1&= &\epsilon_1 \eta_1 + \epsilon_2 \lambda_1, \, 
E_2 = \epsilon_1 \eta_2 +\epsilon_2\lambda_2.
\eea
$\gamma_{1/2}$ and $\epsilon_{1/2}$ are given in [\onlinecite{mathey,mathey2}], 
with 'f' and 'b' replaced by '1' and '2', 
and $\eta_{1/2}$
and $\lambda_{1/2}$ given by:
\bea
\eta_1&=&\sqrt{v_A/\tilde{v}_1}\cos\theta, \, \eta_2=\sqrt{v_A/\tilde{v}_2}\sin\theta,\\
\lambda_1&=&-\sqrt{v_a/\tilde{v}_1}\sin\theta, \,\lambda_2=\sqrt{v_a/\tilde{v}_2}\cos\theta.
\eea
%
%
%



%



\begin{thebibliography}{99}
\bibitem{trap}
Due to the confining trap the transition is, while
 visible, 'blurred' into a gradual cross-over, due to finite size effects, and,
 more importantly, due to the coexistence of several phases in the trap.
 This can also be expected for the phase transitions
 predicted in this paper.
\end{thebibliography}
\end{document}